# Humans incorrectly reject confident accusatory AI judgments

Riccardo Loconte [1,2,*], Merylin Monaro [3,4], Pietro Pietrini[1], Bruno Verschuere [5], Bennett Kleinberg [6]

[1] IMT School of Advanced Studies, Lucca, Italy
[2] Tilburg University, Department of Methodology and Statistics, The Netherlands
[3] University of Padua, Department of General Psychology, Padua, Italy,
[4] University of Padua, Department of Mathematics, Padua, Italy,
[5] University of Amsterdam, Department of Psychology, The Netherlands,
[6] University College London, Department of Security and Crime Science, UK

## Abstract

Automated verbal deception detection using methods from Artificial Intelligence (AI) has been shown to outperform humans in disentangling lies from truths. Research suggests that transparency and interpretability of computational methods tend to increase human acceptance of using AI to support decisions. However, the extent to which humans accept AI judgments for deception detection remains unclear. We experimentally examined how an AI model's accuracy (i.e., the overall performance in deception detection) and confidence (i.e., the model's uncertainty of single-statement prediction) influence human adoption of the model's judgments. Participants (*n*=373) were presented with veracity judgments of an AI model with high or low overall accuracy and various degrees of prediction confidence. The results showed that humans followed predictions from a highly accurate model more than from a less accurate model. Interestingly, the more confident the model, the more people deviated from it, especially if the model predicted deception. We also found that human interaction with algorithmic predictions either worsened the machine's performance or was ineffective. While this human aversion to accept highly confident algorithmic predictions was partly explained by participants' tendency to overestimate their own deception detection abilities, we also discuss how truth-default theory and the social costs of accusing someone of lying help explain the findings.

*Corresponding author: Riccardo Loconte, r.loconte@tilburguniversity.edu

## Introduction

Detecting deception remains one of the most challenging tasks for researchers and practitioners. Traditional methods often rely on subjective cues or limited behavioral indicators (DePaulo et al., 2003; Hartwig & Bond, 2011), making consistent and accurate detection difficult (Bond & DePaulo, 2006a; Hartwig & Bond, 2014). Recent advances in artificial intelligence (AI) offer promising tools to enhance deception detection by analyzing complex patterns in language (Constancio et al., 2023; Hauch et al., 2015), improving both accuracy and scalability.

However, in high-stakes domains, such as in the legal domain, where outcomes directly impact individuals' lives, delegating decisions to machines is strongly disapproved, underscoring the need for including humans in the loop to oversee machine limitations (Kotsoglou & Oswald, 2020; Orsini et al., 2025; van Dijck, 2022). This remarks the need for more research on how humans can benefit from the AI assistance while retaining responsibility for their final decisions and, at the same time, understanding under which conditions AI-based judgments on the credibility of statements are endorsed or rejected by human decision-makers.

### Human deception detection

Decades of research suggest that humans are *truth-biased* and tend to judge statements as truthful more frequently and to be more accurate in detecting truthful statements than deceptive ones. This truth bias is explained by the truth-default theory (TDT; Levine, 2014). According to TDT, people generally assume honesty in communication because deception is quite rare, and its occurrence is limited to a few prolific liars (Levine, 2014; Serota et al., 2010). Indeed, deception is a strategic act that typically occurs only when truthfulness obstructs personal goals (Levine, Kim, & Blair, 2010; Levine, Kim, & Hamel, 2010). As a consequence, people become suspicious only in case of strong triggers that disrupt this default truth assumption (e.g., hidden goals, inconsistencies, or third-party warnings).

Another robust finding is that humans are poor at detecting deception. More specifically, meta-analytical evidence indicates that laypeople's performance in discriminating between deceptive and truthful statements is no better than chance level (Hartwig & Bond, 2011, 2014). Additionally, expertise in the field (e.g., being a police officer dealing with potentially deceptive suspects) does not significantly improve the detection rate (Bond & DePaulo, 2006). In contrast, when relying on evidence-based tools (e.g., Criteria-Based Content Analysis, Reality Monitoring), interviewing strategies (e.g., Strategic Use of Evidence, Imposing cognitive load), and heuristics (e.g., detailedness), human deception detection can be around 70% (Amado et al., 2016; Gancedo et al., 2021; Hartwig et al., 2014; Sporer et al., 2025; Verschuere et al., 2023; Vrij et al., 2008).

However, these traditional deception detection methods require either a one-on-one interaction, especially for strategic interviewing, or human involvement in scoring verbal transcripts for content analysis, making scalability expensive in terms of time and resources (Kleinberg et al., 2019b). Moreover, 70% accuracy is often considered low and unsatisfactory when applied to real-life and sensitive settings (Kleinberg et al., 2019a). A potential avenue for overcoming these limitations lies in the utilization of computational approaches to estimate the credibility of statements.

### Computational approaches

Advancements in computational methods for deception detection have been made possible thanks to progress in computers, machine learning (ML), and natural language processing (NLP), enhancing both scalability and objectivity.

For example, focusing on the well-established knowledge that “lying takes time” (Suchotzki et al., 2017), early computer-aided approaches investigated deception detection through the analysis of mouse (Monaro et al., 2017) and keyboard dynamics (Monaro et al., 2018), while more recent research trained ML models on response latencies and error rates to unexpected questions (Melis et al., 2024). On the other hand, studies focusing on computer vision developed ML models trained at detecting deception through the analysis of facial expressions (see Delmas et al., 2024, for a review of studies).

Among all the computational approaches available, those based on textual data – combining techniques from NLP and ML models - are the most useful because they can be applied to automate manual credibility assessment methods (e.g., Reality Monitoring, Verifiability Approach) and can be used in combination with interviewing techniques. For example, one study employed the proportion of unique named-entities (NER) as a proxy of verifiable details to detect positive and negative deceptive hotel reviews, reaching an area under the curve (AUC) of 0.67 and 0.65, respectively (Kleinberg et al., 2017). Another study relied on a pretrained language model (i.e., Bidirectional Encoder Representations from Transformers - BERT) to detect deceptive utterances in a dataset of transcripts of trial hearings, reaching 71.61% accuracy (Pérez-Rosas et al., 2015). Finally, another work showed that ML algorithms significantly outperformed naïve (accuracy= 54.7%) and expert judges (accuracy= 59.4%) in detecting deception from transcripts of interviews with unexpected questions when trained both on theory-driven (accuracy=69.4%) and data-driven features (accuracy=77.3%) (Loconte et al., 2025).

However, computational approaches for credibility assessment are not yet applied in real-life situations. Previous research has shown that these models show a broad range of accuracy (from 60% to 90%) and have limited generalizability to different domains (Kleinberg et al., 2019; Loconte et al., 2023; Velutharambath & Klinger, 2023). Nevertheless, we might imagine a near future where these limitations might be easily overcome and new language models on deception become available. If this is the case, a key problem with automated verbal deception detection is the inherent difficulty of relying exclusively on AI-based predictions in high-stakes domains. In fact, in sensitive contexts, such as forensic settings, delegating decisions to machines has been strongly disapproved, while including humans in the loop has been encouraged (Kotsoglou & Oswald, 2020; Orsini et al., 2025; van Dijck, 2022). Therefore, further research is needed on how humans can be effectively integrated into the loop to benefit from AI assistance while retaining responsibility. In fact, while hybrid decision-making has already been commonly used in online content moderation (Jhaver et al., 2019)and is becoming more popular in medical diagnoses (Bulten et al., 2020), it remains understudied in the context of deception detection.

## Hybrid decision-making in deception detection

Hybrid decision making consists of integrating human oversight into AI predictions, allowing decision-makers to leverage AI’s analytical strengths and scalability while maintaining responsibility. The rationale behind hybrid decision-making is that AI-based judgments are provided to the human decision-makers to obtain an overall better performance than either mode in isolation.

In the context of deception detection, previous research investigated hybrid decision-making by integrating supervised ML with human judgment and comparing the combined performance to that of each modality in isolation (Kleinberg & Verschuere, 2021). While ML alone achieved a classification accuracy of 69%, human involvement, by fully overruling or adjusting within given boundaries the AI-based predictions, did not significantly improve the deception detection performance and brought the accuracy back to chance level (Kleinberg & Verschuere, 2021). Related work has built on that framework and examined how the availability of a lie-detection algorithm influences human judgment (von Schenk et al., 2024). Participants either received no algorithmic aid (control condition), were always shown AI predictions (forced condition), or could choose to access these predictions (choice condition). Notably, algorithmic availability altered social dynamics: participants were less willing to accuse others without AI support, but those who actively sought and relied on algorithmic predictions (choice condition) were more likely to follow accusatory predictions, compared to those who were passively exposed in the forced condition.

Together, these findings challenge the assumption that human-AI collaboration inherently improves decision-making and underscore the complexity of designing effective hybrid systems. In fact, integrating AI into human decision-making is not merely a technical problem but also a psychological and social challenge. One common limitation in these two studies is that both lacked manipulation of the information provided to participants about the model's overall performance (i.e., accuracy) and uncertainty scores (i.e., confidence) for individual predictions. Indeed, previous studies found accuracy and confidence as key drivers of human trust in AI models. It was found that higher AI stated and observed accuracy reliably increased user reliance (Alufaisan et al., 2021; He et al., 2023; Yin et al., 2019). Similarly, AI confidence levels influenced human self-confidence (Li et al., 2024; Zhang et al., 2020), with alignment effects persisting beyond the interaction (Zhang et al., 2020). This leaves open the question of whether more fine-grained information—such as information on the model's accuracy and prediction confidence— affects hybrid decision-making on deception detection in a more nuanced way. In other terms, we need to understand under which conditions automated judgments on the veracity of statements are endorsed or rejected by human decision-makers, and whether this interplay is beneficial for improved verbal deception detection.

**The present study**

The current study adds to research on hybrid human-machine deception detection by explicitly incorporating and manipulating information about (1) the overall accuracy of the AI model, (2) the AI uncertainty for individual predictions, and (3) the final veracity judgment of the AI model.

To this aim, we developed an experimental task in which participants made veracity judgments about statements presented in contexts relevant to lying and were provided with the predictions of a fictitious AI-based classifier for deception detection.
We conducted a 2 (Accuracy: low=54% vs high=89%) by 5 (Confidence: indecisive, poorly confident, moderately confident, confident, very confident) by 2 (Classification: truthful vs deceptive) mixed-design experiment. The experimental design allows us to isolate the effects of AI model characteristics on human deception judgments when interacting with the model. We hypothesized a two-way interaction between Accuracy and Confidence: for a very accurate model (=89%), humans deviate less from AI predictions when these are made with high confidence compared to low confidence predictions; for a low-accuracy model, confidence does not exert any effect in human deviations, as the underlying predictions are inherently inaccurate regardless of confidence level.

A second research question entails whether information on the model's accuracy and confidence can improve verbal deception detection in the hybrid modality than with the AI mode in isolation. To reply to this research question, we relied on the receiver operating characteristic (ROC) framework to test the diagnostic power of AI- and human AI-assisted judgments and compared the areas under the curve (AUC) to seek potential differences under the two accuracy conditions.

## Materials and Methods

### Ethics and transparency statement

The experiment was designed in accordance with the Declaration of Helsinki and was approved by the local ERB (code: TSB_RP1450). The study design, methods, hypotheses, and confirmatory analyses were preregistered at: https://aspredicted.org/cxy6-t3sq.pdf. All data, materials, and the code to reproduce the analysis are available at https://osf.io/28avp/overview.

### Participants

We recruited 493 participants fluent in English from the general population through Prolific[1]. Each participant provided informed consent. Participants were excluded if they failed at least one of two attention checks. The first one consisted of asking participants to position two sliders at designated points during the task (i.e., -20 and 33). However, since a few participants provided feedback and reported that the slider might have shifted slightly after submission, we were more lenient and retained participants whose responses deviated by only ± 2 points, excluding *n*=66 participants. The second attention check consisted of asking participants, after the task, to recall the model's accuracy in a multiple-choice question with six answer options, excluding *n*=52 participants. After data collection, we checked for the presence of any outliers for the time taken to complete the task. Participants were considered outliers if they completed the task in a time that was two standard deviations lower than the average value, but any outlier was detected. Finally, we checked for careless responding consisting of more than three consecutive identical responses (using the *careless* package in R; Johnson, 2005). Two participants were removed as they provided four and eight consecutive identical responses, respectively. The final remaining sample consisted of 373 participants (52.28% females, 47.72% males) with a mean age of 39.42 years (*SD*=13.61, range: 18-72).

### Study design

The experimental design was 2 (Accuracy: low vs high, between-subjects) by 5 (Confidence: indecisive vs poorly confident vs moderately confident vs confident vs very confident, within-subjects) by 2 (Classification: truthful vs deceptive, within-subjects). Specifically, participants were randomly assigned to one of the two Accuracy conditions (low=54% vs. high=89%). Accuracy conditions were manipulated in the task by showing participants a number of correct predictions equivalent to the model's accuracy rate (e.g., the model with 89% accuracy showed 89% of the time a correct prediction). Additionally, for correct predictions, the veracity and confidence condition of the statement was also kept. In contrast, when a prediction was incorrect, the confidence range and value remained randomly assigned but flipped. Confidence was manipulated within-subjects, with participants evaluating ten statements (half truthful and half deceptive) paired with twice five different confidence levels (i.e., *indecisive, poorly confident, moderately confident, confident, very confident*). Statements were randomly paired with confidence levels and then with a random value from a specific confidence range (for truthful statements: *indecisive* = 0–9, *poorly confident* = 10–19, *moderately confident* = 20–29, *confident* = 30–39, *very confident* = 40–50; for deceptive statements: *indecisive* = 0 – -9; *poorly confident* = -10 – -19; *moderately confident* = -20 – -29; *confident* = -30 – -39; *very confident* = -40 – -50). Classification, i.e., whether the AI model predicted that statement as truthful or deceptive, was inherently manipulated within subjects. Positive values of confidence represented predictions of truthfulness, while negative values represented predictions of deception. Values equal to zero were excluded from the analysis.

[1] An a priori power analysis was conducted using G*Power (Faul et al., 2007) to establish the required number of participants. The results indicated that a sample size of 454 is sufficiently large to achieve a statistical power (1-β) of 0.95 in a mixed design with repeated measures (no. measurements = 10) involving two experimental conditions, given a significance level of α = 0.01, a small effect size (0.1), low correlation between repeated measures (0.1), and a non-sphericity correction ε = 0.1.

## Stimuli

Statements were selected from a subset of a validated dataset of past autobiographical experiences deemed relevant for lying (Loconte et al., 2025). The deception involved in the original dataset was in the form of embedded lies. Specifically, we selected a total of 406 statements (203 truthful and 203 deceptive) about i) being hospitalised and undergoing surgery, ii) being involved in a car accident, iii) causing a car accident, and iv) missing a deadline at work because of bad organisation (see Table 1).

**TABLE 1.** Context for lying, verbatim example of a statement, total number of statements, and average number of words (and SD) per topic.

| Topic | Context for lying | Example statement | No. of statements in topic | Average no. of words (and SD) |
|---|---|---|---|---|
| Being hospitalised and undergoing surgery | The author of the statement might have lied to exaggerate some side effects to receive extra compensation from the health insurance | Six months ago I underwent planned surgery to remove a very large cyst from my scalp and several medium sized cysts. The surgery took longer than planned and as a result I was very unwell afterwards. It took me some time to come round whilst in recovery. The effects of general anaesthetic and anti-emetic drugs caused me to be violently sick, I could not even keep water down. I was rushed to another hospital and admitted to a ward for observation and to administer fluids via a drip. I waited four years for this surgery, during this time my details were lost and several appointments cancelled due to strikes and other issues. If I had received the surgery earlier, the procedure would not have taken so long and I would not have been so unwell after the surgery. | 138 | 157 (95.4) |
| Being involved in a car accident | The author of the statement might have lied to increase the claimed amount of damage received to later get more money from their insurance | Last month, I was in a car accident exiting a mall. The street wasn't all that busy and it was a fairly calm day. I arrived at the four way intersection to a red light and stopped accordingly. When the light turned green, I proceeded to make my turn when a car from the adjacent side ignored their red light and continued passing through and hit my car in the rear. When I exited to confront the individual, they sped away. | 94 | 144 (76.7) |
| Causing a car accident | The author of the statement might have lied to describe the event in a way so that it's not their fault. | A morning i was driving my car to work and i was in a hurry because i was going to late. As i driving i reached traffic lights that was at that moment red. When the traffic lights became green i start moving again and all of a sudden the car in front of me stopped withour a reason, causing me to collide with it. The woman driving the car, came out of the car and she was yelling at me saying that i caued the accident. But the fault was on her, because she stopped suddenly without a reason and all of the other vehicles were moving and also as i noticed she was talking on her cellphone. In my point of view she shouldn't using her phone while driving and be more cautious about the traffic lights and the vehicles behind her. | 30 | 169 (88.3) |
| Missing a deadline at work because of bad organisation | The author of the statement might have lied to find excuses that allow them not to | I missed a timeline of the important work, because of bad organisation of my colleague. I have been trying my best, but they were putting the task on hold since the time has passed and now they are trying to convince everybody that was not their fault, but mine. I have repeated | 144 | 127 (65.5) |

| Topic | Context for lying | Example statement | No. of statements in topic | Average no. of words (and SD) |
|---|---|---|---|---|
| | appear forgetful or disorganised. | many times that the task is important to finish just before 2nd half of May and now there is a lot of to catch up, but they preferred to be passive and take a lot of brakes causing the delay in the job accolated to our team. Now everyone has to work on the task, but our client is losing his patience and judge the whole company by the laziness of the teammates. | | |

The instruction for the main task reads as follows:

> 📖 *In this experiment, you will read twelve short statements about past experiences that are either truthful or lies.*
> *Your task is to guess whether each statement is truthful* ✅ *or a lie* 🤥
> *These statements were randomly selected from a larger dataset containing both truthful and deceptive statements.*
> *To help you with your task, we provide you with the predictions of a lie detection algorithm based on artificial intelligence (AI)* 🤖.
> *You'll see an example on the next pages.*
> *Please note that you should read the statements carefully, as after the task, you will also have to take a quick quiz. The quiz serves to validate your participation.*
>
> *Please read the following statement carefully:*
>
> ***Topic:*** *Being hospitalised and undergoing surgery*
>
> ***Context:*** *They might have lied to exaggerate some side effects and receive extra compensation from the health insurance.*
>
> ***Statement 1****: I was hospitalised for almost three months. I was sick, like really sick and doctors were unable to identify what is the problem. They kept me in hospital for almost a month just to keep on doing blood test and taking samples on my body, until they found out I'm diagnosed with lupus. It was a hard journey, I stated attending Albert Luthuli hospital for Rheumatologist, Ngwelezane hospital for Dermatologist, until I get better.*

The experiment was implemented in a web app designed with the Python-based library *Streamlit* (v.1.45.0; screenshots of the task available here: https://osf.io/28avp/files/4ta5x ).

## Procedure

Participants read twelve statements (i.e., ten experimental statements and two attention checks) and were instructed to 1) act as a judge, 2) read each assigned statement paired with a fictitious AI judgment, and 3) decide whether the statement was truthful or deceptive. The AI judgment was displayed in the form of a slider with a continuous scale from -50 to +50 and 0 as a midpoint. Specifically, values from -50 to 0 represented deceptive statements, and values from 0 to +50 represented truthful statements. Values closer to the extremes of the slider (i.e., -50 or 50) represented very high levels of confidence, and values closer to 0 represented indecisive judgments. This range was intended to mimic class probabilities, which here was defined as *confidence* to ensure naïve understanding. Participants provided their judgment using another slider with the same scale. In this way, we measured participants' judgments and uncertainty in one shot. This similarity across sliders also ensured participants a fair understanding of what confidence means (Figure 1). The slider of each judgment started at the indecisive midpoint of 0, and participants could freely move it in any direction.

**FIGURE 1.** The upper part of the figure depicts the AI judgment in the form of a slider, providing information about the overall veracity judgment plus the confidence level. The lower part of the picture depicts the participant's slider that will be used to provide the final judgment.

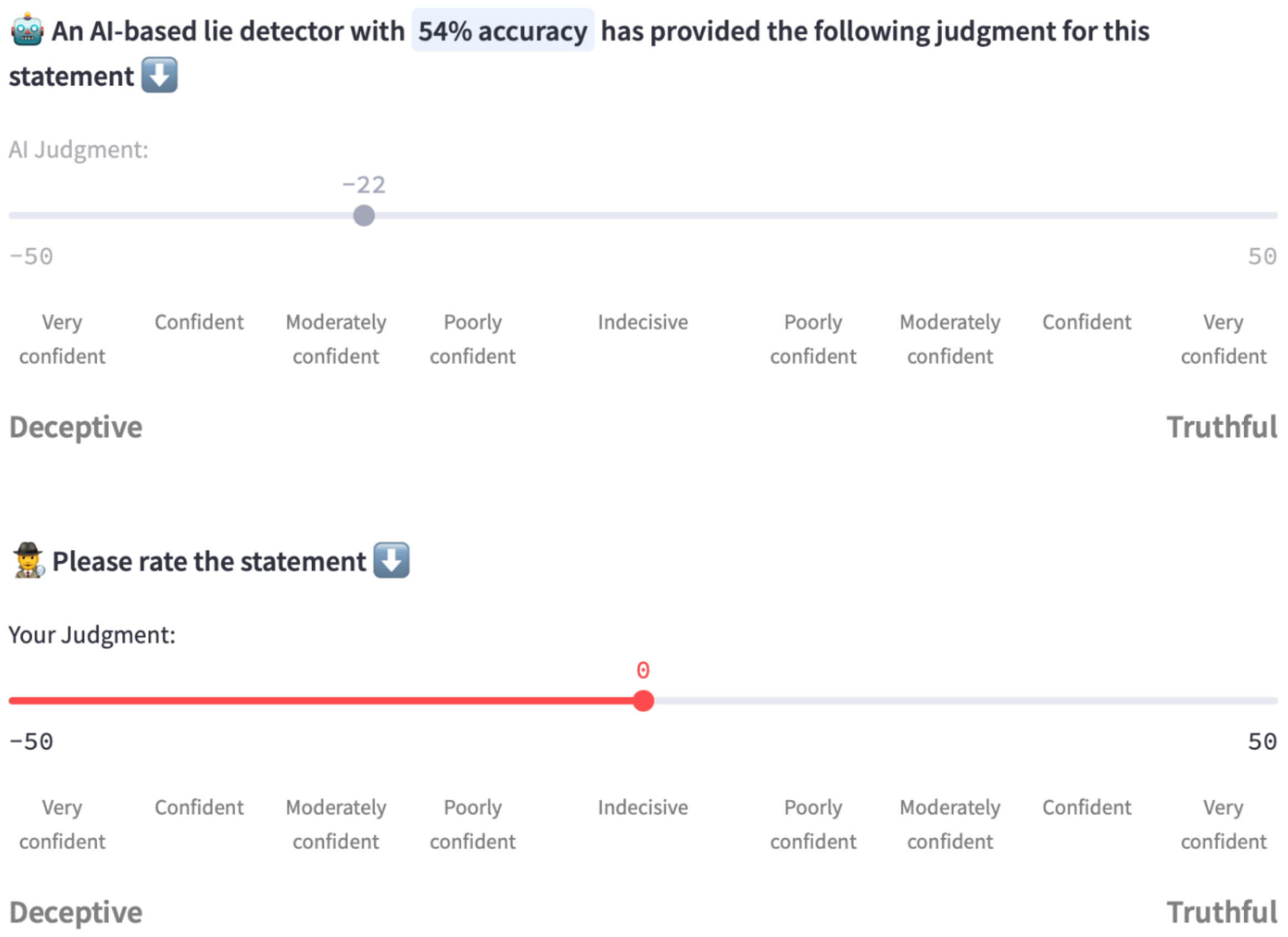

The average duration of each trial (i.e., reading the statement and making a judgment) was 49.03 seconds (*SD*=28.97, *Median*=43.72). After performing the task, participants rated on a 10-point scale their level of motivation (1=Not at all,10=Very much), difficulty of the task (1=Very easy, 10=Very difficult), familiarity with machine learning models (1=Not familiar at all, 10=Very familiar), and their performance in the deception detection task (1=Algorithm performance is better, 10=My performance is better) as well as that of the average human compared to the AI model (1=Algorithm performance is better, 10=Human performance is better). After the experiment ended, participants were thanked and debriefed about the fictitious nature of the AI model. Participants were remunerated with GBP 1.80. The duration of the task was, on average, 21.46 minutes (*SD*=51.94, *Median*=15.82).

## Analysis plan

The main and preregistered analyses are reported in the *Confirmatory analyses* paragraph of the *Results* section. For the main analysis, we rescaled the judgments from -50 - +50 to 0-100 to have all values in the positive range. The dependent variable was computed as the difference between human judgment and AI judgment ($\Delta y = y_{human} - y_{AI}$), where positive values indicated that humans judged the statement as more truthful than the AI, and negative values indicated that humans judged the statement as more deceptive. We used a linear mixed model to account for the nested structure of the data, as each participant provided multiple judgments and each statement was judged multiple times. We determined the optimal model structure by comparing three nested linear mixed-effects models in predicting Δy (see Supplementary Materials, SM 1). Based on the results, Δy was modelled as a function of the main effects and full interactions of Accuracy (low = 54% vs. high = 89%), Confidence (five levels: indecisive, poorly confident, moderately confident, confident, very confident), and Classification (truthful vs. deceptive) as fixed effects. The model included one random intercept for participants and one for statements, to account for individual- and stimulus-related differences in baseline deviation.

As a robustness check, we re-run the same analysis by including as covariates participants' scores on i) motivation, ii) difficulty of the task, iii) familiarity with ML models, and iv) goodness of their performance and that of the v) average human compared to AI. All analyses were conducted in R, using the *lme4* and *lmerTest* libraries. The alpha level for statistical significance was 0.01, and effect sizes are reported with 99% confidence intervals.

**Exploratory analysis**

Given that one-fourth ($n$=120) of the sample was excluded from the analysis due to failing attention and statistical checks, we re-run our main analysis on the whole sample as a robustness check.

Moreover, we also analyzed the data from a different perspective by examining absolute values of human deviations ($|\Delta y| = |y_{AI} - y_{human}|$), namely deviations from model predictions irrespective of whether participants judged statements as more truthful or more deceptive. Put differently, the analysis focused on the magnitude of the deviation rather than the direction of the deviation.

Finally, since findings show that participants tend to deviate from AI predictions under specific conditions, we tested whether these human corrections improve detection between deceptive and truthful statements. With this aim, we used the ROC framework, which evaluates the diagnostic power of a continuous response variable (here: the human judgment) by plotting sensitivity against 1-specificity across all possible thresholds. The area under the curve (AUC) quantifies this performance, ranging from 0 (=perfectly inaccurate) to 1 (=perfectly accurate), with 0.50 indicating chance-level discrimination. Differences in AUC between conditions were tested using the DeLong test (DeLong et al., 1988).

# Results

## Preliminary analysis

Descriptive statistics of participants' deviation from AI scores (Δy) across conditions are reported in Table S1.

A permutation *t*-test with 9,999 permutations indicated no difference between the high vs low accuracy condition in i) the motivation to perform the task well, ii) the perceived difficulty of the task, and iii) the level of ML familiarity (Table 2). However, participants in the low-accuracy condition scored significantly higher on believing that AI performs worse than the average human compared to those in the high-accuracy condition, *d*=0.31 [99% CI: 0.04, 0.58]. Participants in the low-accuracy condition also scored significantly higher on believing that their own judgments were better than AI judgments compared to the high accuracy condition, *d*=0.29 [0.02, 0.56].

**TABLE 2.** Mean, standard deviation, and Cohen's *d* of the difference between covariates across accuracy conditions.

| | Accuracy low | | Accuracy high | | |
|---|---|---|---|---|---|
| | Mean | *SD* | Mean | *SD* | *d* (99% CI) |
| Motivation | 8.75 | 1.43 | 8.66 | 1.53 | 0.06 (-0.21; 0.33) |
| Difficulty | 3.65 | 2.79 | 3.55 | 2.83 | 0.03 (-0.23; 0.30) |
| ML familiarity | 6.11 | 2.47 | 6.01 | 2.47 | 0.04 (-0.23, 0.31) |
| AI vs Average Human | 6.67 | 1.84 | 6.06 | 2.14 | 0.31 (0.04, 0.58) * |
| AI vs Yourself | 6.51 | 1.81 | 5.95 | 2.01 | 0.29 (0.02, 0.56) * |

*Abbreviations:* SD = *Standard deviation;* CI = Confidence intervals.
* p < .01

After computing participant-level averages for deviations (Δy) and absolute deviations (|Δy|), associations with self-reported covariates using Spearman rank correlations were examined. Results indicated a small positive correlation between the average deviation (Δy) from AI predictions and participants' belief that the average human performs better than the AI model ($r_s$(371)=0.18, *p*=0.002). Small positive correlations were also found between the average absolute deviation (|Δy|) and participants' perception of outperforming the model ($r_s$(371)=0.34, *p*<0.001) and individuals' beliefs that the average human performs better than the AI model ($r_s$(371)=.29, *p*<0.001). Correlations between the remaining covariates (i.e., motivation, difficulty, and ML familiarity) were not found to be significant (see Figure S1 in SM 2 for the correlation matrix). This suggests that participants who perceived the average human, or themselves, as more capable in spotting deception than the AI model tended to display larger deviations from the algorithmic predictions.

## Confirmatory analyses

For the main analysis, a linear mixed-effects model was fitted with random intercepts for participants and statements (see Table S2 in SM3). Using the *anova( )* function, type III F-tests were conducted to investigate the main and interaction effects of the categorical predictors.

The 2 (Accuracy: low vs. high) by 5 (Confidence: indecisive vs poorly confident vs moderately confident vs confident vs very confident) by 2 (Classification: truthful vs deceptive) ANOVA showed significant main effects of Classification, a significant two-way interaction effect of Accuracy by Classification and Confidence by Classification, and a significant three-way interaction effect of Accuracy by Confidence by Classification (Table 3).

**TABLE 3.** Type III Analysis of Variance for the Linear Mixed Model predicting human deviation.

| Effect | Sum of Sq | Mean Sq | Num DF | Den DF | *F*-value | $\eta^2$ (99% CI) |
|---|---|---|---|---|---|---|
| Accuracy | 9 | 9 | 1 | 377.5 | 0.02 | 0.00 (0.00, 0.00) |
| Confidence | 928 | 232 | 4 | 180.0 | 0.46 | 0.01 (0.00, 0.05) |
| Classification | 514,132 | 514,132 | 1 | 3,425.6 | **1026.21**** | 0.23 (0.20, 0.26) |
| Accuracy × Confidence | 2,594 | 649 | 4 | 3,241.6 | 1.29 | 0.00 (0.00, 0.01) |
| Accuracy × Classification | 20,655 | 20,655 | 1 | 3,426.1 | **41.23**** | 0.01 (0.00, 0.02) |
| Confidence × Classification | 287,802 | 71,951 | 4 | 3,426.4 | **143.61**** | 0.14 (0.12, 0.017) |
| Accuracy × Confidence × Classification | 8,370 | 2,092 | 4 | 3,426.4 | **4.18*** | 0.01 (0.00, 0.01) |

*Note*. For this model, the bobyqa optimiser was used to fit the model. Results are based on Satterthwaite's approximation for degrees of freedom. In **bold** are reported significant *F*-values.
Abbreviations: Num DF= number of degrees of freedom; Sq = squares
* $p < .01$, ** $p < .001$

The main effect of Classification suggested that the human deviation from model predictions was significantly higher when the model predicted deceptive (M=19.40, SE=0.76) than when it predicted truthful (M=-5.25, SE=0.77), d=0.55 [0.50, 0.59].

For the two-way interaction of Accuracy by Classification, the human deviation from the model's predictions of deceptiveness was significantly higher for the low-accuracy model (M=21.80, SE=0.97) than for the high-accuracy one (M=16.99, SE=0.99), d=0.13 [0.04, 0.22], moving the AI judgment towards more truthful but with a greater extent for the low-accuracy model. Similarly, the average deviation from the model's predictions of truthfulness was significantly more pronounced in the low-accuracy condition (M=-7.79, SE=0.98) than in the high-accuracy condition (M=-2.71, SE=1.02), d=-0.13 [-0.22, -0.05]. The negative deviation values reflect participants changing the initial AI judgment towards more deceptive, but did so to a greater extent in the high-accuracy than in the low-accuracy.

For the two-way interaction of Confidence by Classification, human deviation from AI predictions of deceptiveness increased significantly with each step up in the confidence range, except between the indecisive and poorly confident, and poorly confident and moderately confident range, suggesting that the higher the model's confidence, the higher the human deviation from AI judgments of deception. Similarly, for AI predictions of truthfulness, deviations were significantly more pronounced across steps up in the confidence range, except between indecisive and poorly confident, and between moderately confident and confident (see Table S3 for pairwise contrasts in SM).

Most importantly, we found a significant three-way interaction between Accuracy, Confidence, and Classification, which we unpacked by Accuracy (Figure 2).

**FIGURE 2.** Average values of deviation Δy across Accuracy (low vs. high), Confidence (indecisive vs. poorly confident vs. moderately confident vs. confident vs. very confident), and Classification (deceptive vs. truthful) conditions.

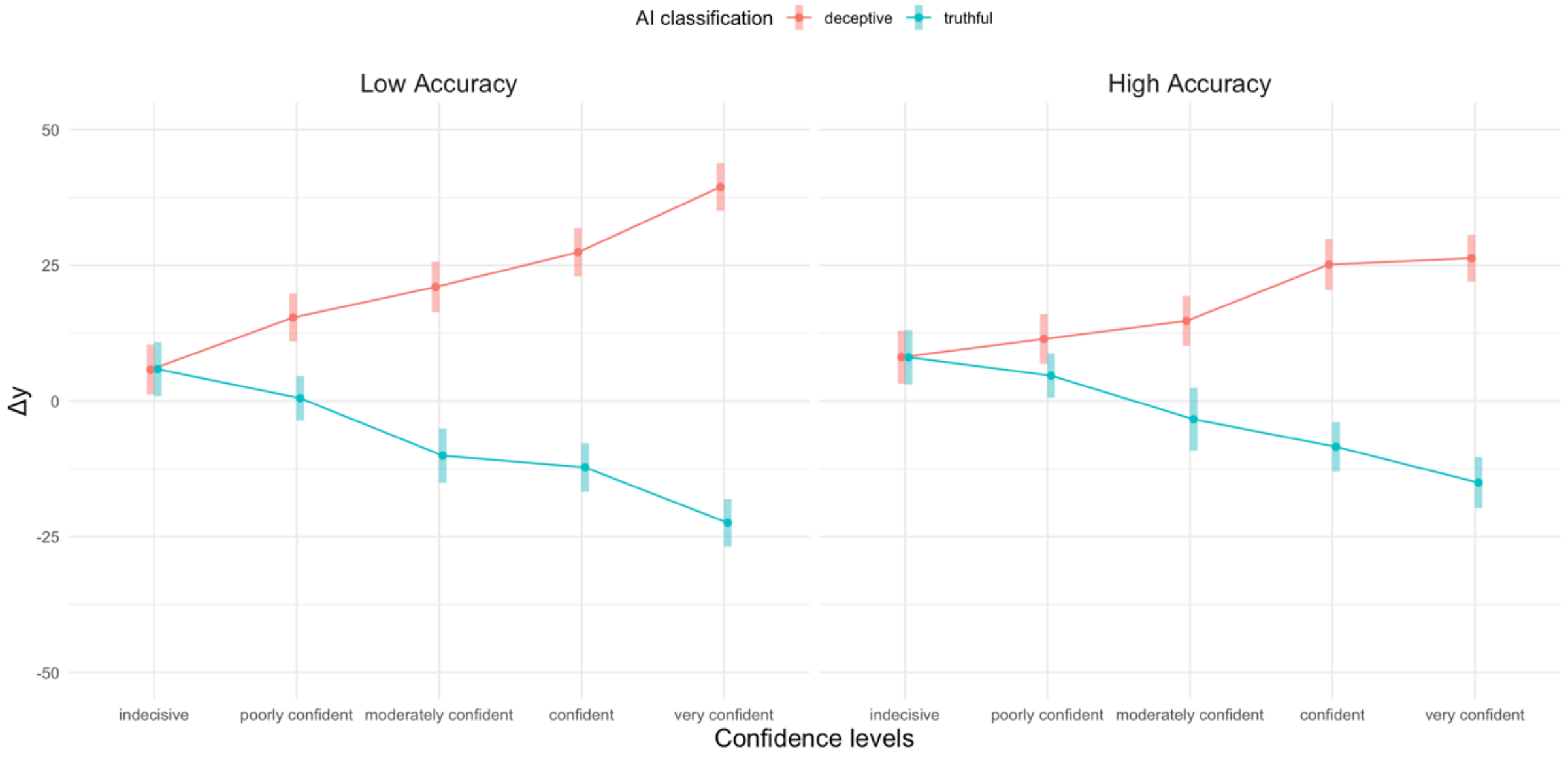

*Note.* Error bars represent 99% CI.

### *Low-accuracy model*

Under the low-accuracy condition, the Confidence by Classification ANOVA indicated a significant main effect of Classification and a significant interaction effect. For the main effect of Classification, $F(1, 1812.3)=696.31$, $p<.001$, $\eta^2=0.28$ [0.23, 0.32], human deviation was significantly higher for AI predictions of deceptiveness ($M=21.96$, $SE=1.03$) than for predictions of truthfulness ($M=-7.73$, $SE=1.05$), $d=0.62$ [0.55, 0.69], indicating that humans tended to disagree more from AI accusations of deception.

The significant interaction between Confidence and Classification, $F(4, 1832.56)=84.25$, $p<0.001$, $\eta^2=0.16$ [0.12, 0.19]. revealed that human deviation from AI predictions of deceptiveness increased significantly with each step up in the confidence range, except between *poorly confident* and *moderately confident*, and between *moderately confident* and *confident* levels. For AI predictions of truthfulness, deviation was more pronounced within steps up in confidence levels, except between *indecisive* and *poorly confident*, and between *moderately confident* and *confident*. These findings suggest that, for both types of AI predictions (i.e., truthful and deceptive), human deviation from the initial AI judgment tends to grow as the AI expresses greater confidence (see Table S4 for pairwise contrasts in SM), although in different directions, namely judging deceptive statements as more truthful and truthful statements as more deceptive.

### *High-accuracy model*

Similarly to the low accuracy condition, the Confidence by Classification ANOVA indicated a significant main effect of Classification and a significant interaction effect for the high-accuracy condition. Concerning Classification, F(1, 1605.6)=364.58, p<.001, $\eta$2=0.19 [0.14, 0.23], humans deviated more from AI predictions of deceptiveness (M=17.15, SE=0.95) than from predictions of truthfulness (M=-2.79, SE=0.98), d=0.48 [0.41, 0.54].

The interaction between Confidence and Classification, $F(4, 1611.0)=58.76$, $p<.001$, $\eta^2=0.13$ [0.09, 0.17] revealed a more nuanced pattern of human deviation across confidence levels. When the AI classified statements as deceptive, deviations increased significantly only at higher levels of confidence (i.e., *confident* and *very confident*) compared to lower levels. Yet, the difference between *confident* and *very confident* was not significant. In contrast, when the AI classified statements as truthful, deviations remained more constant, with significant differences emerging at least every two steps up the confidence scale (e.g., *moderately confident* was significantly different from *very confident* but not from *confident*). These results suggest that, under high-accuracy conditions, only high levels of confidence significantly influence human deviations from AI predictions of deceptiveness, while deviations from truthful AI predictions increase more gradually across confidence levels (see Table S5 in SM).

## Exploratory analysis

### *Robustness check*

As a robustness check, the same analysis was re-run on a new model that included participants' scores of covariates and on a new model including the participants who failed the attention checks ($n$=120) and were previously excluded from analysis. Our findings were unaffected by the inclusion of the covariates (see Table S6). However, when previously excluded participants were included in the analysis, the three-way interaction effect was no longer significant (see Table S7). This suggests that, while the observed effects were not driven by individual differences in the covariates, participants' attention during the task was important.

### *Magnitude of deviation*

Full analyses of the magnitude of human deviation (i.e., irrespective of whether participants judged statements as more truthful or more deceptive) are provided in SM6, while a summary of important results is reported below.

A first relevant finding was that, regardless of the condition, the magnitude of deviation was always significantly greater than zero (see Figure 3). Moreover, we observed a significant interaction between AI classification and confidence level, $F(4, 3463.4)=13.13$, $p<.001$, $\eta^2=0.01$ [0.01, 0.03]. As illustrated in Figure 3 (left panel), deviations from model predictions of truthfulness did not vary significantly across confidence levels (see Table S11). In contrast, participants' magnitude of deviations significantly increased with increases in confidence levels for predictions of deceptiveness (see Table S11).

Finally, we found a significant interaction between confidence level and model accuracy, $F(4, 3293.6)=5.27$, $p<.001$, $\eta^2=0.006$ [0.00, 0.01] (see Figure 3, right panel). Between the two conditions, a significant difference in magnitude was observed only for *very confident* predictions stemming from the low-accuracy model than from the high-accuracy one. Relatedly, no differences were found for the high-accuracy condition within confidence levels, while, for the low-accuracy condition, the average magnitude at the *very confident* level was significantly greater than those at the remaining lower levels (see Table S12).

This pattern of findings confirms that participants tend to reject more over-confident predictions, especially those stemming from a low-accuracy model and those predicting deception.

**FIGURE 3.** Average values of absolute deviation between i) Classification and Confidence (left panel) and ii) Accuracy and Confidence (right panel).

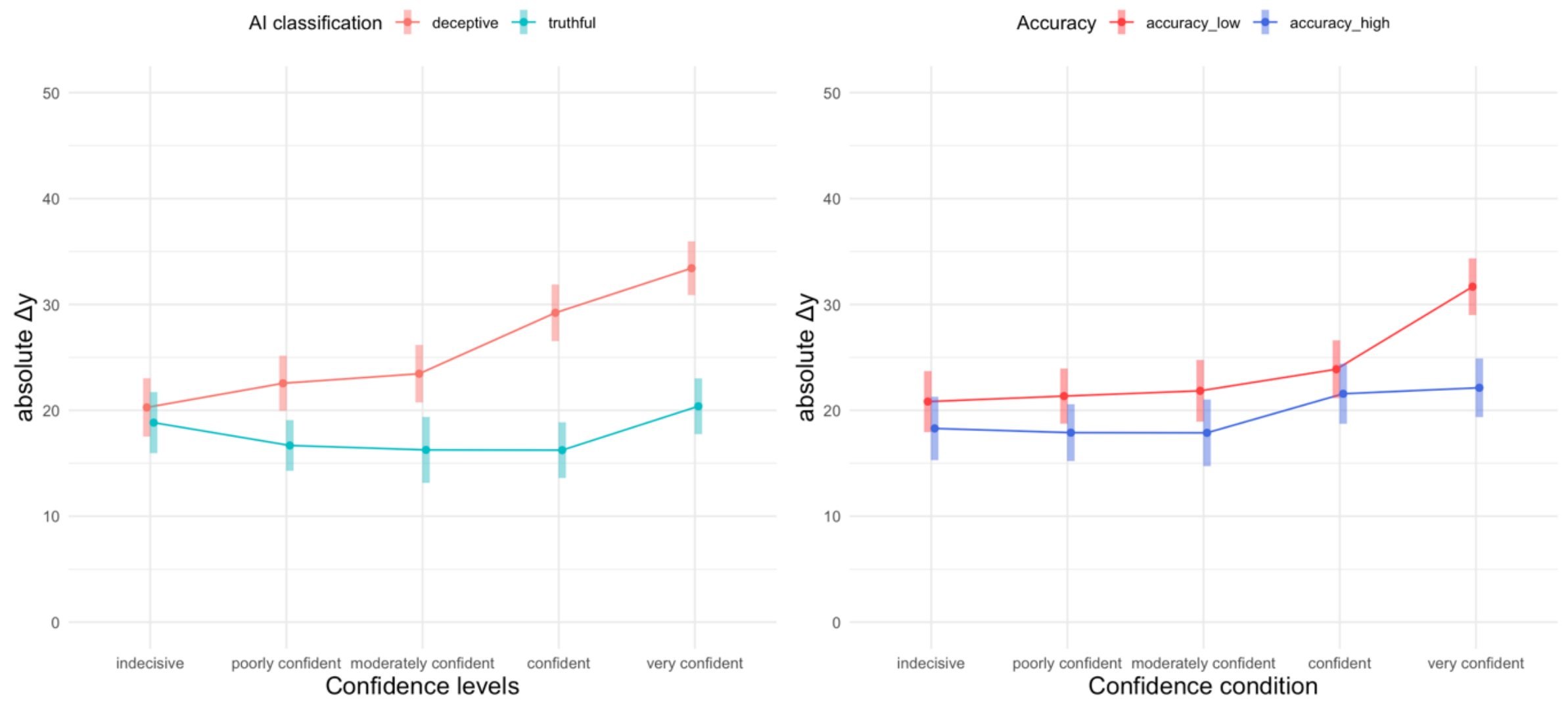

*Note.* Error bars represent 99% CI.

### ***Human performance in detecting deception***

Participants guided by the high-accuracy model achieved a detection performance that was significantly higher than chance (AUC=0.76 [0.73; 0.79]) and significantly higher than participants' performance under the low-accuracy model condition (AUC=0.57 [0.54; 0.60]; DeLong test: $D(3692.5)=11.23$, $p<.001$). Yet, this detection rate was significantly lower than that of the highly accurate model (AUC=0.90 [0.88; 0.92]; DeLong test for two correlated ROC curves: $z=-12.42$, $p<.001$), suggesting that human interaction with the model's predictions worsens detection performance compared to relying entirely on the AI model.

As for participants' performance under the low-accuracy model condition (AUC=0.57 [0.54;0.60]), this was not significantly better than the model's performance alone (AUC=0.54 [0.51,0.57]), $z=-1.82$, $p=.06$.

This pattern of findings suggests that participants were influenced by model predictions in both conditions, with human performance mirroring that of the corresponding model, yet without overcoming the model's detection rate.

## Discussion

This study investigated hybrid deception detection by exploring under which conditions algorithmic judgments on statement veracity are endorsed or rejected by human decision-makers. Specifically, this study focused on the role of the model's accuracy and uncertainty for individual predictions. Using statements about past events deemed relevant for lying (Loconte & Kleinberg, 2025), participants were provided with predictions from a fictitious AI-based deception classifier and were instructed to act as a judge and evaluate the veracity of statements. The experimental task was designed to isolate the effects of the model's characteristics (i.e., accuracy and confidence) on human deviations from the model's judgments. We predicted a two-way interaction effect between the model's accuracy and confidence, so that, only for the highly accurate model, but not the lowly accurate one, human deviation significantly decreases as the model's confidence increases. In other words, for a very accurate model, higher confidence in predictions produces greater endorsement of the model's predictions by humans. However, this hypothesis was only partially supported.

### Main findings

Contrary to our expectations, humans tended to reject AI judgments more when these were made with high confidence. That rejection was more pronounced when the prediction indicated *deception,* compared to predictions indicating *truthfulness,* and when stemming from a low-accuracy model than from a very accurate one. As for the direction of the deviation, the more the AI was confident in predicting *deception*, the more participants leaned their judgment towards less deceptive (hence, as more truthful). In contrast, the more the model predicted *truthfully* with greater confidence, the more participants leaned towards less truthful (hence, more deceptive).

These findings are partially in contrast with previous studies showing that accuracy and confidence are key drivers of human trust, with higher levels of AI accuracy and confidence being associated with higher levels of trust in AI models (Alufaisan et al., 2021; He et al., 2023; Li et al., 2024, 2025; Yin et al., 2019). While our findings are consistent with existing literature on AI-accuracy, suggesting that an increase in the stated model accuracy leads to a greater degree of human trust in these models (He et al., 2023; Yin et al., 2019), we observed a different direction for the effect of confidence (Li et al., 2024, 2025). Specifically, we observed greater deviations from *very confident* predictions than from low-confident ones. Furthermore, we did not replicate a full truth-bias effect that was found in previous studies on hybrid deception detection (Kleinberg & Verschuere, 2021; Levine, 2014; von Schenk et al., 2024). While participants tended to judge deceptive statements as more truthful, especially when predicted with higher levels of confidence, this was not mirrored for truthful statements, which were judged as less truthful under the same high-confidence conditions in our experiment.
Finally, our findings add to the emerging work on hybrid deception detection. Others report that AI availability can increase the accusation rate in deception settings (von Schenk et al., 2024). In our work, we show that providing users with transparent information about the model's performance and uncertainty can discourage them from blindly following an accusatory AI judgment. Confidence cues, in particular, play a critical role in this sense: when users learn that an algorithm displays high confidence in a deceptive judgment, they are less likely to accept it.

We argue that there are two possible explanations for this high-confidence aversion effect. First, humans' beliefs about their own ability in detecting deception, or that of the average human, could explain why participants tended to disagree more with AI predictions. Recent studies have suggested that the type of task assigned to an AI influences the extent to which humans rely on its output (Thorp et al., 2025): when performing objective tasks (e.g., counting dots in an image), people conform more to AI, while, in subjective tasks (e.g., interpreting emotions in an abstract image), they align more with responses of other humans (Riva et al., 2022). Being deception detection somehow subjective, it is likely that humans tend to rely more on their judgment than on that of an AI model. Furthermore, by overestimating human ability in spotting deception, individuals may perceive the high-confident model's judgments as overconfidence, thus raising skepticism toward AI predictions. Prior research already showed that laypeople and practitioners who strongly believed in the wrong cues for deception detection (e.g., behavioral and non-verbal cues such as gaze aversion) were also less inclined to rely on the most relevant ones (i.e., verbal cues). Similarly, individuals who hold strong beliefs about the average human performance in detecting deception might be less inclined to trust an AI model, regardless of its stated level of accuracy.

The distrust in AI possibly also explains why we did not replicate a full truth-bias effect, as participants who relied more on their capabilities might have disagreed with AI judgments regardless of the classification label. However, since these associations are small, other factors may contribute to explaining these findings. A second explanation, through the lens of Truth Default Theory (TDT; Levine, 2014), is that participants were more comfortable making judgments with moderate levels of confidence. According to TDT, people generally assume honesty unless they have reasons to suspect otherwise. Highly confident AI accusations of deception may have lacked sufficient justification to prevail over this truth-default status, thus favoring a more cautious stance, more aligned with their expectation of truthfulness, and acknowledging the social costs of falsely accusing someone of lying.

**Humans-AI performance in deception detection**

This study also tested whether human oversight on machine predictions was beneficial or simply worsened the algorithmic performance. We found participants exposed to predictions from a low-accuracy model performed only slightly better than chance (accuracy: 0.57), but without significantly improving the accuracy of the model itself. In contrast, participants guided by the high-accuracy model, despite showing a detection rate that was significantly better than chance (accuracy: 0.76), performed significantly lower than the high-accuracy model (accuracy: 0.90). Notably, had participants under the latter condition relied entirely on the algorithmic predictions, their performance would have closely matched the 89% accuracy. Moreover, this drop suggests that participants' modifications were not just a mere deviation in the confidence level but represent a substantial modification in the classification itself. In other words, participants deviated from AI judgments to an extent that reversed the classification judgment from *deceptive* to *truthful*, and vice versa.

These findings align with previous research suggesting that the human interplay with algorithmic predictions either worsens machine performance or is simply ineffective (Kleinberg & Verschuere, 2021; von Schenk et al., 2024). What we added to the previous body of research is that adding a transparency and interpretability layer on the model's accuracy and confidence is not sufficient to achieve a hybrid-performance that is better than that of the model in isolation. One possible interpretation for this finding is that the nature of the task itself might be a critical factor. Prior meta-analytical evidence showed that human-AI collaboration tends to be beneficial in tasks where humans already outperform AI, whereas impairments are observed when AI alone outperforms humans (Vaccaro et al., 2024). Given that humans are known to be poor at deception detection (Bond & DePaulo, 2006b; DePaulo et al., 2003; Hartwig & Bond, 2011), it is plausible that, for this task, a hybrid modality is unlikely to exceed the performance of an AI model in isolation.

**Limitations and future research**

A few limitations are worth mentioning for this study. First, our study entails the fact that the model's predictions were always available to participants. In related work (von Schenk et al., 2024), some participants always received predictions from a lie-detection algorithm (i.e., forced condition) before making their own judgments, while others could choose to request such predictions (i.e., choice condition), and others performed the task without any AI aid (i.e., control condition). Their findings show that model availability increases the accusation rate, especially among those who actively seek such algorithmic predictions. Our study adds more nuance to these findings by showing that including a transparency layer on the model's performance and uncertainty can discourage users from blindly following an accusatory AI judgment. However, unlike von Schenck et al. (2024), our study explored the role of transparency on trust exclusively under a *forced condition*, namely, with participants always having access to the AI predictions. Future studies should, instead, investigate the role of transparency in human trust by also including conditions where participants are free to choose whether to access or not to AI predictions (i.e., choice condition).
Second, we run our experiment in a non-incentivised task, and we did not provide any information about possible consequences that the author of the statements would have experienced if accused of lying. It might be that by adding this information, participants could have been even more averse towards over-confident AI predictions. Future research may try to investigate the role of accuracy and confidence in an incentivized task that also incorporates the possibility to choose whether or not participants want to seek AI predictions.
Third, in our study, humans were not aware of their inherent limited capability for deception detection, and, indeed, we found that greater deviations positively correlated with this erroneous belief that their own performance, or that of the average human, is usually better than that of an AI model. Future research should explore whether and how informing users about their inherent limitations in deception detection changes their trust in AI predictions. Finally, while we informed participants on the model's accuracy and predicting confidence, the role of the model's explainability was not explored in this study. Explainability has been proposed as a mechanism to improve trust and decision-making by providing users with insights into the rationale behind predictions and by potentially enabling better understanding and error correction (Alufaisan et al., 2021). However, while some studies from other domains support the benefits of explainability (Balasubramaniam et al., 2023; Oswald et al., 2018; Sadeghi et al., 2024), meta-analytic evidence suggests that its impact on performance may not exceed that of AI predictions alone (Schemmer et al., 2022). These mixed findings highlight the need for further research into how and whether explainability can improve human trust in AI, specifically for deception detection.

## Conclusion

This study investigated how the accuracy and confidence of a fictitious AI model for deception detection influence human reliance on AI-based veracity judgments. Human judges refrained from relying on highly confident AI predictions, especially when accusing someone of deception or stemming from a low-accuracy model. Human aversion toward AI predictions was partially explained by participants' tendency to overestimate their own deception detection capabilities. In line with the truth-default theory, such aversion also reflected users' cautiousness in deception accusations, given the social costs of falsely accusing someone of lying. This caution, however, comes at the expense of reducing individuals' own ability to detect deception effectively.

# Supplementary Material

## 1. Optimal linear-mixed model structure

To determine the optimal model structure, we compared three nested linear mixed-effects models predicting the absolute deviation from AI judgments.

Model 1 included fixed effects for the two-way interaction between Accuracy and Confidence, considered AI Classification as a covariate, and included a random intercept for participants and a random intercept for statements. Model 2 added a three-way interaction among the predictors. Model 3 included the same fixed effects as Model 2 but added random slopes for Confidence by participant to account for individual differences in how participants adjust their deviation across confidence levels.

1) $\Delta y = Accuracy * Confidence + Classification + (1 \mid \text{Participant_id}) + (1 \mid \text{Statement_id})$
2) $\Delta y = Accuracy * Confidence * Classification + (1 \mid \text{Participant_id})\ ) + (1 \mid \text{Statement_id})$
3) $\Delta y = Accuracy * Confidence * Classification + (1 + \text{Confidence} \mid \text{Participant_id}) +$
$(1 \mid \text{Statement_id})$

A likelihood ratio test comparing the models (refitted using maximum likelihood) showed that Model 2 provided a significantly better fit than Model 1, $\chi^2(9) = 589.41$, $p < .001$, and Model 3 did not significantly outperform Model 2, $\chi^2\ (14) = 12.35$, $p = .578$.

Based on these results, we selected Model 2 as the final model, indicating that the inclusion of a three-way interaction and a random intercept for participants and statements was enough to significantly improve the model fit.

## 2. Preliminary Analysis

**TABLE S1.** Mean, standard deviation, and median values of participants' deviation from AI (Δy) across conditions

| | | Accuracy low | | | Accuracy high | | |
|---|---|---|---|---|---|---|---|
| Classification | Confidence | *M* | *SD* | *Median* | *M* | *SD* | *Median* |
| Deceptive | Indecisive | 5.83 | 24.75 | 8 | 7.85 | 21.80 | 6 |
| | Poorly Confident | 15.87 | 26.31 | 15.5 | 11.57 | 23.51 | 6 |
| | Moderately Confident | 21.01 | 25.82 | 15 | 14.9 | 23.43 | 10 |
| | Confident | 27.21 | 28.07 | 20 | 25.16 | 26.15 | 19.5 |
| | Very Confident | 39.5 | 29.90 | 31 | 26.37 | 26.45 | 20 |
| Truthful | Indecisive | 6.07 | 23.09 | 12 | 8.20 | 19.54 | 9.5 |
| | Poorly Confident | 0.12 | 22.55 | 6 | 4.55 | 18.91 | 9 |
| | Moderately Confident | -10.15 | 22.06 | -3 | -3.68 | 20.37 | 1.5 |
| | Confident | -12.07 | 22.73 | -6 | -8.45 | 22.0 | -1.5 |
| | Very Confident | -22.61 | 25.17 | -15 | -15.06 | 20.88 | -9 |

**FIGURE S1.** Correlation matrix of Spearman correlations between the participant's average deviation and absolute deviation and covariates of motivation, difficulty, machine learning familiarity, belief in performing better than the AI, and belief that the average human performs better than the AI.

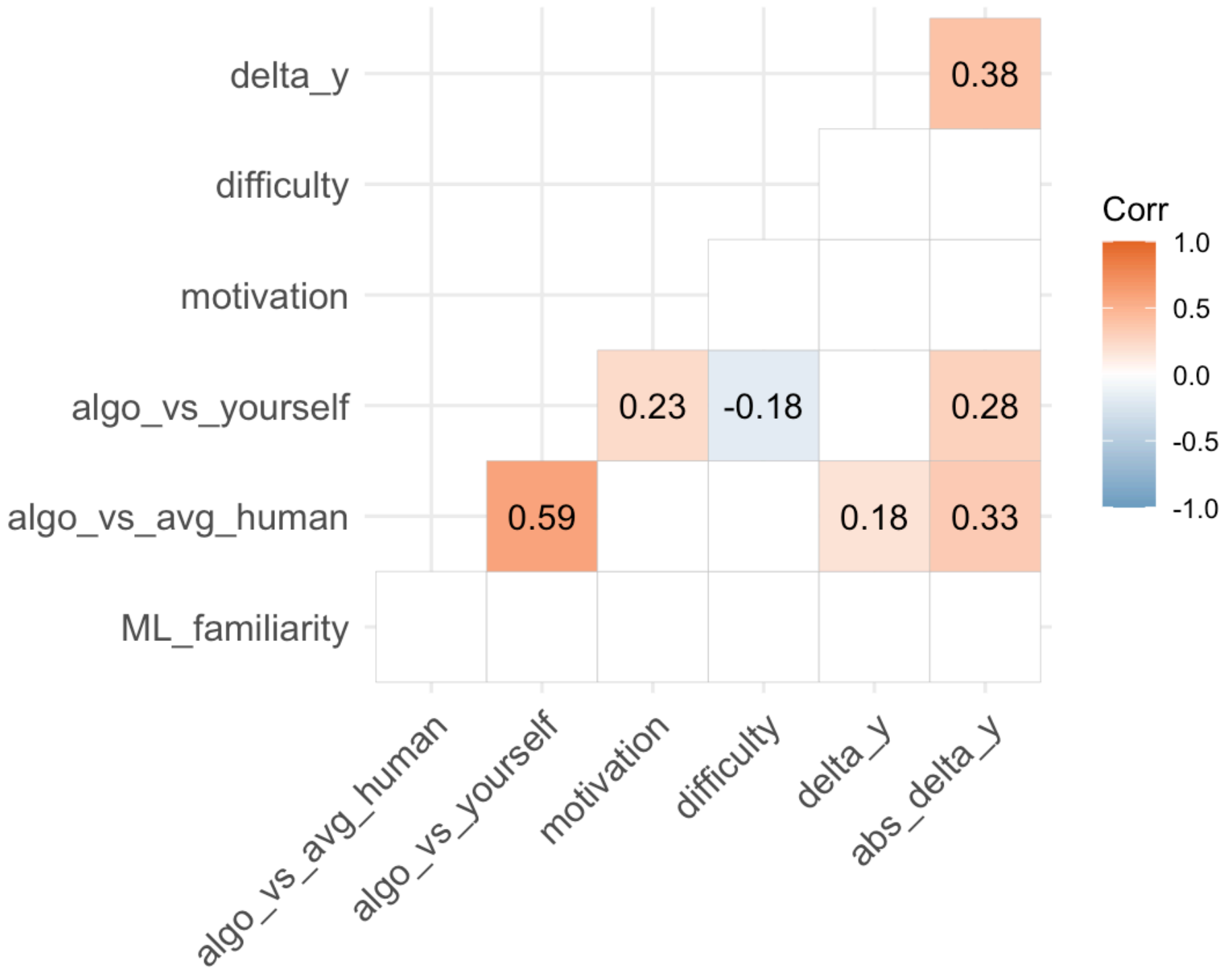

*Note:* Only significant correlations at $p < .01$ after *false discovery rate* correction for multiple comparisons are displayed in the correlation matrix.
*Abbreviation:*
abs_delta_y = average absolute human deviation at the participant-level (dependent variable)
delta_y = average human deviation at the participant-level (dependent variable)
algo_vs_avg_human = covariate of belief that the average human's performance is better than that of the AI
algo_vs_yourself = covariate of belief that an individual's performance is better than that of the AI
difficulty = covariate of the difficulty of the task
ML_familiarity = covariate of familiarity with machine learning models
Motivation = covariate of motivation in performing the task well

## 3. Confirmatory Analysis

**TABLE S2.** Variance, Standard Deviation (SD), and adjusted Intraclass Correlation Coefficient (ICC) for Random Effects from Model 2.

| Random Effects | | | |
|---|---|---|---|
| Group | Variance | SD | ICC |
| Participant_id | 39.44 | 6.28 | 0.07 |
| Statement_id | 35.91 | 5.99 | 0.06 |
| Residual | 501.00 | 22.38 | - |

**TABLE S3.** Pairwise contrasts for Confidence range by Classification.

| Pairwise contrast | AI Classification = deceptive | | AI Classification = truthful | |
|---|---|---|---|---|
| | *Diff* (*SE*) | *d* (99% CI) | *Diff* (*SE*) | *d* (99% CI) |
| indecisive – poorly confident | -6.36 (2.14) | -0.14 (-0.27, -0.02) | 3.86 (2.13) | 0.09 -(0.04, 0.22) |
| indecisive – moderately confident | -11.14 (2.27) ** | -0.27 (-0.42, -0.13) | 13.10 (2.43) ** | 0.27 (0.14, 0.40) |
| indecisive – confident | -19.01 (2.21) ** | -0.45 (-0.58, -0.31) | 16.84 (2.23) ** | 0.38 (0.25, 0.52) |
| indecisive – very confident | -26.30 (2.15) ** | -0.63 (-0.78, -0.49) | 24.93 (2.22) ** | 0.55 (0.42, 0.69) |
| poorly confident – moderately confident | -4.78 (2.18) | -0.12 (-0.26, -0.02) | 9.23 (2.25) ** | 0.21 (0.08, 0.34) |
| poorly confident – confident | -12.65 (2.11) ** | -0.30 (-0.43, -0.17) | 12.98 (2.03) ** | 0.34 (0.20, 0.48) |
| poorly confident – very confident | -19.95 (2.05) ** | -0.48 (-0.61, -0.35) | 21.06 (2.01) ** | 0.53 (0.39, 0.67) |
| moderately confident – confident | -7.87 (2.24) ** | -0.20 (-0.34, -0.05) | 3.74 (2.35) | 0.08 (-0.05, 0.21) |
| moderately confident – very confident | -15.16 (2.18) ** | -0.39 (-0.54, -0.24) | 11.83 (2.33) ** | 0.25 (0.12, 0.38) |
| confident – very confident | -7.29 (2.11) * | -0.18 (-0.31, -0.04) | 8.09 (2.12) ** | 0.20 (0.06, 0.33) |

*Note. M* = estimated marginal means; *SE* = standard error; *d* = Cohen's *d*, *CI* = confidence interval.
* *p* < .011
** *p* < .001

**TABLE S4.** Pairwise contrasts for Confidence range by Classification under the low-accuracy condition.

| Low-accuracy condition | | | | |
|---|---|---|---|---|
| Pairwise contrast | AI Classification = deceptive | | AI Classification = truthful | |
| | *Diff* (*SE*) | *d* (99% CI) | *Diff* (*SE*) | *d* (99% CI) |
| indecisive – poorly confident | -9.60 (2.90) * | -0.16 (-0.28, -0.03) | 4.37 (2.92) | 0.07 (-0.05, 0.19) |
| indecisive – moderately confident | -15.47 (3.06) ** | -0.27 (-0.41, -0.13) | 14.62 (3.24) ** | 0.22 (0.09, 0.35) |
| indecisive – confident | -20.79 (2.96) ** | -0.35 (-0.49, -0.22) | 17.12 (3.06) ** | 0.27 (0.14, 0.40) |
| indecisive – very confident | -33.44 (2.91) ** | -0.57 (-0.70, -0.43) | 26.83 (3.02) ** | 0.42 (0.30, 0.55) |
| poorly confident – moderately confident | -5.86 (2.96) | -0.10 (-0.23, 0.03) | 10.24 (2.98) * | 0.17 (0.04, 0.31) |
| poorly confident – confident | -11.19 (2.85) * | -0.19 (-0.31, -0.06) | 12.75 (2.78) ** | 0.23 (0.10, 0.36) |
| poorly confident – very confident | -23.84 (2.81) ** | -0.40 (-0.52, -0.27) | 22.46 (2.73) ** | 0.40 (0.27, 0.54) |
| moderately confident – confident | -5.32 (3.01) | -0.09 (-0.23, 0.04) | 2.50 (3.11) | 0.04 (-0.09, 0.17) |
| moderately confident – very confident | -17.97 (2.97) ** | -0.32 (-0.46, -0.18) | 12.21 (3.07) ** | 0.20 (0.07, 0.33) |
| confident – very confident | -12.65 (2.87) ** | -0.22 (-0.35, -0.09) | 9.71 (2.88) * | 0.17 (0.04, 0.30) |

*Note. Diff* = estimated difference of marginal means; *SE* = standard error; *d* = Cohen's *d*, *CI* = confidence interval.
* *p* < .01
** *p* < .001

**TABLE S5.** Pairwise contrasts for Confidence range by Classification under the high-accuracy condition.

| High-accuracy condition | | | | |
|---|---|---|---|---|
| Pairwise contrast | AI Classification = deceptive | | AI Classification = truthful | |
| | *Diff* (*SE*) | *d* (99% CI) | *Diff* (*SE*) | *d* (99% CI) |
| indecisive – poorly confident | -3.42 (2.68) | -0.06 (-0.18, 0.06) | 3.31 (2.60) | 0.06 (-0.06, 0.19) |
| indecisive – moderately confident | -6.75 (2.77) | -0.13 (-0.27, 0.01) | 11.30 (3.06) * | 0.17 (0.05, 0.29) |
| indecisive – confident | -17.21 (2.74) ** | -0.31 (-0.44, -0.18) | 16.42 (2.72) ** | 0.30 (0.17, 0.43) |
| indecisive – very confident | -18.69 (2.64) ** | -0.36 (-0.49, -0.22) | 22.61 (2.74) ** | 0.38 (0.26, 0.51) |
| poorly confident – moderately confident | -3.32 (2.65) | -0.07 (-0.20, 0.07) | 7.99 (2.85) | 0.13 (0.01, 0.25) |
| poorly confident – confident | -13.79 (2.62) ** | -0.25 (-0.37, -0.13) | 13.12 (2.47) ** | 0.27 (0.14, 0.41) |
| poorly confident – very confident | -15.27 (2.51) ** | -0.29 (-0.42, -0.17) | 19.30 (2.49) ** | 0.38 (0.25, 0.51) |
| moderately confident – confident | -10.46 (2.72) * | -0.21 (-0.35, -0.07) | 5.12 (2.95) | 0.08 (-0.04, 0.20) |
| moderately confident – very confident | -11.95 (2.61) ** | -0.26 (-0.41, -0.11) | 11.31 (2.97) * | 0.17 (0.05, 0.29) |
| confident – very confident | -1.48 (2.58) | -0.03 (-0.16, 0.10) | 6.19 (2.62) | 0.12 (-0.01, 0.24) |

*Note. Diff* = estimated difference of marginal means; *SE* = standard error; *d* = Cohen's *d*, *CI* = confidence interval.
* *p* < .01
** *p* < .001

## 4. Robustness check by including covariates

Covariates:

- Motivation: "*How much were you motivated to perform well?*" (0=Not at all, 10=Very much)
- Difficulty: "*How difficult did you find the study?*" (0=Very easy, 10=Very difficult)
- ML familiarity: "*How familiar are you with AI-based algorithms?*" (0=Not familiar at all, 5=Neutral, 10=Very familiar)
- AI vs Average human: "*How good do you think the **average human performance** is compared to the performance of the AI-based lie detector in predicting whether a statement is true or false?*" (0=Algorithm's performance is better, 5=Equal, 10=Human's performance is better)
- AI vs Yourself: "*How good do you think **your performance** is compared to the performance of the AI-based lie detector in predicting whether a statement is true or false?*" (0=Algorithm's performance is better, 5=Equal, 10=My performance is better)

**TABLE S6.** Type III Analysis of Variance for the Linear Mixed Model predicting human deviation.

| Effect | Covariates | Sum of Sq | Mean Sq | Num DF | Den DF | *F*-value | $\eta^2$ (99% CI) |
|---|---|---|---|---|---|---|---|
| Accuracy | Excluded | 9 | 9 | 1 | 377.5 | 0.02 | 0.00 (0.00, 0.00) |
| | Included | 244 | 244 | 1 | 373.0 | 0.49 | 0.00 (0.00, 0.03) |
| Confidence | Excluded | 928 | 232 | 4 | 180.0 | 0.46 | 0.01 (0.00, 0.05) |
| | Included | 920 | 230 | 4 | 180.3 | 0.46 | 0.01 (0.00, 0.05) |
| Classification | Excluded | 514,132 | 514,132 | 1 | 3,425.6 | **1026.21**** | 0.23 (0.20, 0.26) |
| | Included | 512,845 | 512,845 | 1 | 3,427.3 | **1023.20**** | 0.23 (0.20, 0.26) |
| Accuracy × Confidence | Excluded | 2,594 | 649 | 4 | 3,241.6 | 1.29 | 0.00 (0.00, 0.01) |
| | Included | 2,669 | 667 | 4 | 3,242.5 | 1.22 | 0.00 (0.00, 0.01) |
| Accuracy × Classification | Excluded | 20,655 | 20,655 | 1 | 3,426.1 | **41.23**** | 0.01 (0.00, 0.02) |
| | Included | 20,854 | 20,854 | 1 | 3,431.6 | **41.61**** | 0.01 (0.00, 0.02) |
| Confidence × Classification | Excluded | 287,802 | 71,951 | 4 | 3,426.4 | **143.61**** | 0.14 (0.12, 0.017) |
| | Included | 287,726 | 71,931 | 4 | 3,460.5 | **143.51**** | 0.14 (0.12, 0.017) |
| Accuracy × Confidence × Classification | Excluded | 8,370 | 2,092 | 4 | 3,426.4 | **4.18*** | 0.01 (0.00, 0.01) |
| | Included | 8,470 | 2,118 | 4 | 3,468.6 | **4.22*** | 0.01 (0.00, 0.01) |

*Note*. For this model, the bobyqa optimiser was used to fit the model. Results are based on Satterthwaite's. In **bold** are reported significant *F*-values.
Abbreviations: Num DF= number of degrees of freedom; Sq = squares
* $p < .001$

## 5. Robustness check by including previously excluded participants

**TABLE S7.** Type III Analysis of Variance for the Linear Mixed Model predicting human deviation.

| Effect | Sample | Sum of Sq | Mean Sq | Num DF | Den DF | *F*-value | $\eta^2$ (99% CI) |
|---|---|---|---|---|---|---|---|
| Accuracy | Reduced | 9 | 9 | 1 | 377.5 | 0.02 | 0.00 (0.00, 0.00) |
| | Full | 53 | 53 | 1 | 499.7 | 0.10 | 0.00 (0.00, 0.00) |
| Confidence | Reduced | 928 | 232 | 4 | 180.0 | 0.46 | 0.01 (0.00, 0.05) |
| | Full | 1730 | 432 | 4 | 182.2 | 0.85 | 0.02 (0.00, 0.08) |
| Classification | Reduced | 514,132 | 514,132 | 1 | 3,425.6 | **1026.21**** | 0.23 (0.20, 0.26) |
| | Full | 749,367 | 749,367 | 1 | 4,591.6 | **1470.48**** | 0.24 (0.22, 0.27) |
| Accuracy × Confidence | Reduced | 2,594 | 649 | 4 | 3,241.6 | 1.29 | 0.00 (0.00, 0.01) |
| | Full | 1,942 | 486 | 4 | 4,341.4 | 0.95 | 0.00 (0.00, 0.00) |
| Accuracy × Classification | Reduced | 20,655 | 20,655 | 1 | 3,426.1 | **41.23**** | 0.01 (0.00, 0.02) |
| | Full | 13,602 | 13,602 | 1 | 4,596.6 | **26.69**** | 0.01 (0.00, 0.01) |
| Confidence × Classification | Reduced | 287,802 | 71,951 | 4 | 3,426.4 | **143.61**** | 0.14 (0.12, 0.017) |
| | Full | 393,315 | 98,329 | 4 | 4,624.5 | **192.95**** | 0.14 (0.12, 0.017) |
| Accuracy × Confidence × Classification | Reduced | 8,370 | 2,092 | 4 | 3,426.4 | **4.18*** | 0.01 (0.00, 0.01) |
| | Full | 5,866 | 1,466 | 4 | 4,627.2 | 2.88 | 0.00 (0.00, 0.01) |

*Note*. For this model, the bobyqa optimiser was used to fit the model. Results are based on Satterthwaite's. *Reduced* sample refers to the sample after excluding participants who failed attention and statistical checks. *Full* sample refers to the whole sample of participants collected without any exclusion. In **bold** are reported significant *F*-values. Abbreviations: Num DF= number of degrees of freedom; Sq = squares
* $p < .01$
** $p < .001$

## 6. Absolute values of human deviation

Model equation: $|\Delta y| = Accuracy * Confidence * Classification + (1|\text{ Participant_id}) ) + (1|\text{ Statement_id})$

**TABLE S8.** Variance, Standard Deviation (SD), and adjusted Intraclass Correlation Coefficient (ICC) for Random Effects.

| Random Effects | | | |
|---|---|---|---|
| Group | Variance | SD | ICC |
| Participant_id | 45.84 | 6.77 | 0.13 |
| Statement_id | 2.05 | 1.43 | 0.01 |
| Residual | 318.11 | 17.84 | - |

**TABLE S9.** Type III Analysis of Variance for the Linear Mixed Model predicting absolute values of human deviation.

| Effect | Sum of Sq | Mean Sq | Num DF | Den DF | *F*-value | $\eta^2$ (99% CI) |
|---|---|---|---|---|---|---|
| Accuracy | 7,126 | 7,126 | 1 | 376.9 | **22.40**** | 0.06 (0.01, 0.13) |
| Confidence | 27,204 | 6,801 | 4 | 159.4 | **21.38**** | 0.35 (0.19, 0.47) |
| Classification | 56,292 | 56,292 | 1 | 3,440.4 | **176.96**** | 0.05 (0.03, 0.07) |
| Accuracy × Confidence | 6,712 | 1,678 | 4 | 3,293.6 | **5.27**** | 0.00 (0.00, 0.01) |
| Accuracy × Classification | 1,401 | 1,401 | 1 | 3,441.9 | 4.40 | 0.00 (0.00, 0.01) |
| Confidence × Classification | 16,704 | 4,1761 | 4 | 3,463.4 | **13.13**** | 0.01 (0.01, 0.03) |
| Accuracy × Confidence × Classification | 591 | 148 | 4 | 3,470.4 | 0.46 | 0.00 (0.00, 0.00) |

*Note*. For this model, the bobyqa optimiser was used to fit the model. Results are based on Satterthwaite's approximation for degrees of freedom. In **bold** are reported significant *F*-values.
Abbreviations: Num DF= number of degrees of freedom; Sq = squares
* $p < .01$, ** $p < .001$

**TABLE S10.** Pairwise contrasts for the main effects of accuracy, confidence, and classification.

| Main effect | Pairwise contrast | *Diff.* (*SE*) | *d* (99% CI) |
|---|---|---|---|
| Accuracy | low-accuracy – high-accuracy | 4.36 (0.92) ** | 0.24 (0.11, 0.38) |
| Confidence | indecisive – poorly confident | -0.06 (0.99) | -0.00 (-0.19, 0.18) |
| | indecisive – moderately confident | -0.30 (1.07) | -0.02 (-0.23, 0.18) |
| | indecisive – confident | -3.17 (1.02) | -0.24 (-0.44, -0.04) |
| | indecisive – very confident | -7.35 (1.01) ** | -0.55 (-0.75, -0.34) |
| | poorly confident – moderately confident | -0.24 (1.01) | -0.02 (-0.22, 0.18) |
| | poorly confident – confident | -3.11 (0.95) | -0.25 (-0.45, -0.05) |
| | poorly confident – very confident | -7.28 (0.94) ** | -0.58 (-0.78, -0.37) |
| | moderately confident – confident | -2.86 (1.04) | -0.23 (-0.44, -0.01) |
| | moderately confident – very confident | -7.04 (1.03) ** | -0.55 (-0.77, -0.33) |
| | confident – very confident | -4.18 (0.97) ** | -0.33 (-0.54, -.013) |
| Classification | deceptive – truthful | 8.1 (0.61) ** | 0.23 (0.18, 0.27) |

Abbreviations: *Diff.* = Estimated difference of marginal means; *SE* = standard error; *d* = Cohen's d; CI = confidence interval
** $p < .001$

**TABLE S11.** Pairwise contrasts for Confidence range by Classification.

| Pairwise contrast | AI Classification = deceptive | | AI Classification = truthful | |
|---|---|---|---|---|
| | *Diff* (*SE*) | *d* (99% CI) | *Diff* (*SE*) | *d* (99% CI) |
| indecisive – poorly confident | -2.28 (1.38) | -0.07 (-0.17, 0.04) | 2.16 (1.36) | 0.06 (-0.04, 0.17) |
| indecisive – moderately confident | -3.18 (1.41) | -0.11 (-0.23, 0.02) | 2.58 (1.57) | 0.07 (-0.04, 0.17) |
| indecisive – confident | -8.94 (1.40) ** | -0.27 (-0.39, -0.16) | 2.60 (1.43) | 0.08 (-0.03, 0.18) |
| indecisive – very confident | -13.15 (1.36) ** | -0.41 (-0.53, -0.30) | -1.55 (1.43) | -0.04 (-0.14, 0.06) |
| poorly confident – moderately confident | -0.90 (1.37) | -0.03 (-0.15, 0.09) | 0.42 (1.44) | 0.01 (-0.09, 0.12) |
| poorly confident – confident | -6.65 (1.36) ** | -0.20 (-0.30, -0.09) | 0.44 (1.28) | 0.02 (-0.10, 0.13) |
| poorly confident – very confident | -10.87 (1.32) ** | -0.33 (-0.43, -0.22) | -3.70 (1.29) | -0.12 (-0.22, -0.01) |
| moderately confident – confident | -5.75 (1.39) ** | -0.20 (-0.32, -0.07) | 0.02 (1.50) | 0.00 (-0.11, 0.11) |
| moderately confident – very confident | -9.96 (1.35) ** | -0.36 (-0.48, -0.23) | -4.12 (1.50) | -0.11 (-0.22, -0.01) |
| confident – very confident | -4.21 (1.34) | -0.13 (-0.24, -0.02) | -4.15 (1.35) | -0.13 (-0.23, -0.02) |

*Note. Diff* = estimated difference of marginal means; *SE* = standard error; *d* = Cohen's *d*, *CI* = confidence interval.
** $p < .001$

**TABLE S12.** Pairwise contrasts for Confidence range by Accuracy condition.

| Pairwise contrast | Low-accuracy condition | | High-accuracy condition | |
|---|---|---|---|---|
| | *Diff* (*SE*) | *d* (99% CI) | *Diff* (*SE*) | *d* (99% CI) |
| indecisive – poorly confident | -0.52 (1.34) | -0.02 (-0.12, 0.09) | 0.40 (1.38) | 0.01 (-0.09, 0.11) |
| indecisive – moderately confident | -1.02 (1.43) | -0.03 (-0.15, 0.09) | 0.41 (1.52) | 0.01 (-0.10, 0.12) |
| indecisive – confident | -3.06 (1.37) | -0.10 (-0.21, 0.02) | -3.28 (1.43) | -0.09 (-0.20, 0.01) |
| indecisive – very confident | -10.85 (1.36) ** | -0.34 (-0.46, -0.23) | -3.84 (1.41) | -0.11 (-0.21, -0.01) |
| poorly confident – moderately confident | -0.50 (1.35) | -0.02 (-0.14, 0.10) | 0.02 (1.44) | 0.00 (-0.11, 0.11) |
| poorly confident – confident | -2.54 (1.29) | -0.09 (-0.20, 0.03) | -3.67 (1.33) | -0.11 (-0.22, -0.01) |
| poorly confident – very confident | -10.33 (1.27) ** | -0.35 (-0.46, -0.23) | -4.24 (1.31) | -0.13 (-0.24, -0.03) |
| moderately confident – confident | -2.04 (1.38) | -0.07 (-0.19, 0.05) | -3.69 (1.47) | -0.11 (-0.22, 0.00) |
| moderately confident – very confident | -9.83 (1.37) ** | -0.34 (-0.47, -0.22) | -4.26 (1.46) | -0.12 (-0.23, -0.01) |
| confident – very confident | -7.79 (1.31) ** | -0.27 (-0.38, -0.15) | -0.57 (1.36) | -0.02 (-0.13, 0.09) |

*Note. Diff* = estimated difference of marginal means; *SE* = standard error; *d* = Cohen's *d*, *CI* = confidence interval.
** $p < .001$